# Mutual Redundancies in Inter-human Communication Systems:

## Steps Towards a Calculus of Processing Meaning

Loet Leydesdorff [i]* and Inga A. Ivanova [ii]

**Abstract**
The study of inter-human communication requires a more complex framework than Shannon's (1948) mathematical theory of communication because "information" is defined in the latter case as meaningless uncertainty. Assuming that meaning cannot be communicated, we extend Shannon's theory by defining mutual redundancy as a *positional* counterpart of the *relational* communication of information. Mutual redundancy indicates the surplus of meanings that can be provided to the exchanges in reflexive communications. The information is redundant because based on "pure sets," that is, without subtraction of mutual information in the overlaps. We show that in the three-dimensional case (e.g., of a Triple Helix of university-industry-government relations), mutual redundancy is equal to mutual information ($R_{xyz} = T_{xyz}$); but when the dimensionality is even, the sign is different. We generalize to the measurement in *N* dimensions and proceed to the interpretation. Using Luhmann's social-systems theory and/or Giddens' structuration theory, mutual redundancy can be provided with an interpretation in the sociological case: different meaning-processing structures code and decode with other algorithms. A surplus of ("absent") options can then be generated that add to the redundancy. Luhmann's "functional (sub)systems" of expectations or Giddens' "rule-resource sets" are positioned mutually, but coupled operationally in events or "instantiated" in actions. Shannon-type information is generated by the mediation, but the "structures" are (re-)positioned towards one another as sets of (potentially counterfactual) expectations. The structural differences among the coding and decoding algorithms provide a source of additional options in reflexive and anticipatory communications.

[i] Amsterdam School of Communication Research (ASCoR), University of Amsterdam, Kloveniersburgwal 48, 1012 CX Amsterdam, The Netherlands; loet@leydesdorff.net ; * corresponding author.
[ii] Department of International Education & Department of Economics and production management, Far Eastern Federal University, Office 514, 56 Aleutskaya st., Vladivostok 690950, Russia; Phone: 007 (423) 2436080, fax: 007 (423) 2457200; inga.iva@mail.ru.

**Introduction**

This study originated from the puzzle caused by the possible change in sign when mutual information is measured in more than two dimensions. Mutual information in more than two dimensions has been considered as a *signed* information measure (Yeung, 2008), but it has remained poorly understood given the Shannon framework that allows only for forward transmission of information in messages and the positive generation of probabilistic entropy (Krippendorff, 2009a and b). In the discussions about extending the number of helices beyond a Triple Helix of university-industry-government relations to quadruple, quintuple, or n-tuple sets of helices, the problem of possible sign changes in this operationalization had remained unresolved (Leydesdorff & Sun, 2009; Kwon *et al.*, 2012; cf. Carayannis & Campbell, 2009 and 2010; Leydesdorff, 2012a).

When the number of possible options by cross-tabling the different codes of communication in the various helices increases, the uncertainty relative to the maximum information content can decrease. This possible reduction of uncertainty ("synergy"; Leydesdorff & Strand, in press) in the case of an increasing number of options has sociological relevance since both uncertainty and expectations are involved in inter-human communications. How can the rich domain of inter-human—reflexive—communications be modeled so that the methodological apparatus of Shannon's (1948) mathematical theory of communication can fruitfully be applied? (e.g., Krippendorff, 1986; Theil, 1972)

**Theoretical background**

In the final chapter of his study entitled "Toward a Sociological Theory of Information," Harold Garfinkel ([1952], 2008)—the founding father of ethnomethodology—listed no fewer than 92 "Problems and Theorems" that were at the time awaiting elaboration at the interface between sociology and information theory. Like others (e.g., his teacher Talcott Parsons), Garfinkel was acutely aware that on the basis of the work of the American pragmatists (Herbert Mead, William James) and European phenomenologists (Edmund Husserl, Alfred Schütz), the study of "large scale self-regulating systems" (p. 208) was placed on the agenda of sociology by then-recent advances in information theory by Von Neumann & Morgenstern (1944), Shannon & Weaver (1949), Deutsch (1950), Miller (1951), Ruesch & Bateson (1951), and others.

Most authors in the sociological tradition work within a paradigm where information cannot be considered content-free and meaningless, as it is defined in Shannon's (1948) mathematical theory of communication. Shannon defined information as "uncertainty"—operationalized as probabilistic entropy—and not as "informative" in the sense of reducing uncertainty.[1] His coauthor and commentator, Warren Weaver, called this definition "bizarre," but "potentially so penetratingly clearing the air that one is now, perhaps for the first time, ready for a real theory of meaning" (at p. 27). However, a theoretical impasse was generated that would last for decades (cf. MacKay, 1969): on the one side, Shannon's summations of probabilities and their logarithms cannot add up to "a difference that makes a difference" for a system of reference (e.g., an observer; Bateson, 1972: 315), and, on the other, social scientists are not inclined to abandon "action" or "an observer" as the system of reference.

---

[1] Varela (1979, at p. 266) noted the original etymology of "in-formare" (Horn & Wilburn, 2005).

Hayles (1990) compared these diametrically opposed definitions of information—as uncertainty or as reduction of uncertainty—to defining a glass as half-full or half-empty. She mentions Weaver's (1949: 26) proposal to introduce the communication of "meaning" as "another box in the diagram which, inserted between the information source and the transmitter, would be labeled 'semantic noise,' the box previously labeled as simply 'noise' now being labeled 'engineering noise' " (1990: 193; see Figure 1).

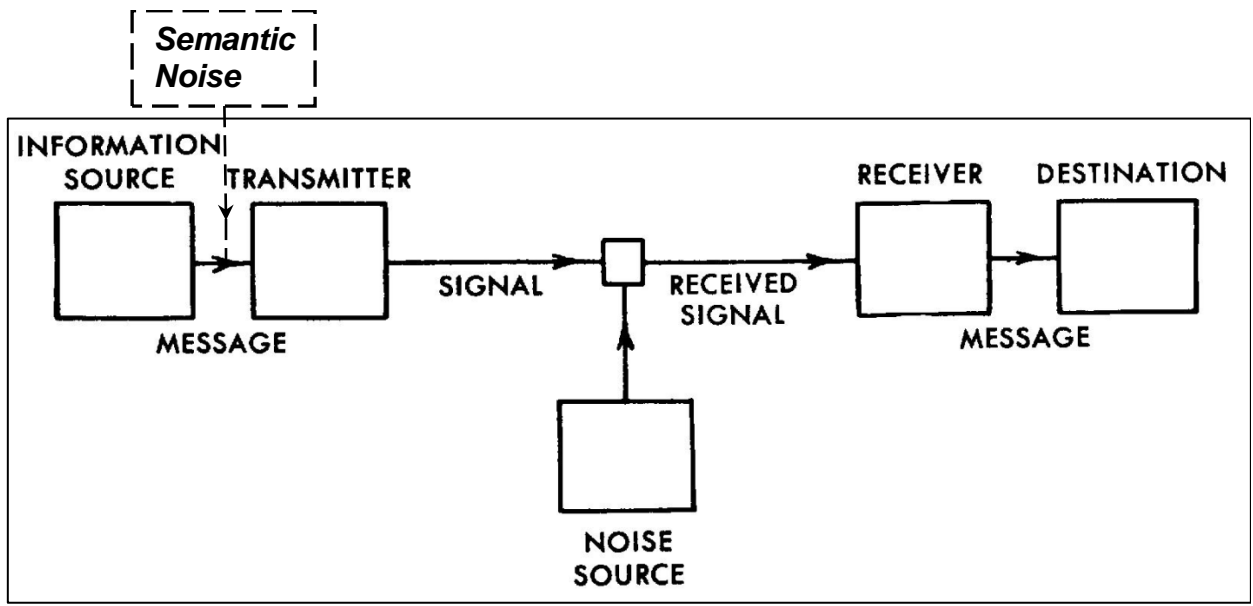

**Figure 1**: Schematic diagram of a general communication system. Source: Shannon (1948, at p. 380); with Weaver's box of "semantic noise" added.

The communication of (Shannon-type) information necessarily adds uncertainty because Shannon (1948) defined information as probabilistic entropy: $H = -\sum_i p_i * \log_2 p_i$. Probabilistic entropy is coupled to Gibbs' formula for thermodynamic entropy $S = k_B * H$. In Gibbs' equation, $k_B$ is the Boltzmann constant that provides the dimensionality Joule/Kelvin to *S*, while *H* is dimensionless and can be measured in bits of information (or any other base for the logarithm) given a probability distribution (containing uncertainty). The Second Law of thermodynamics states that entropy increases with each operation, and Shannon-type information is therefore always positive (because $k_B$ is a constant).

Unlike information that is transmitted from a sender to a receiver in accordance with the arrow of time, meaning is provided from the perspective of hindsight, that is, against the arrow of time. In other words, meaning operates incursively (Dubois, 1998)—that is, reflexively with reference to the present state—whereas information is generated recursively—that is, with reference to a previous state (e.g., as a sender for a receiver). A local reversal of the time axis within a system is compatible with the Second Law if the entropy increases at the aggregated level of the total system (e.g., Wicken, 1989). Local pockets of negative entropy (e.g., niches) are thus possible within the entropy flow.

In summary, Weaver's "semantic noise" may add more to the redundancy than to the uncertainty when the communication is reflexive. In Weaver's formulation (at p. 13): "This [redundancy] is the fraction of the structure of the message which is determined not by the free choice of the sender, but rather by the accepted statistical rules governing the use of the symbols in question." From this perspective, semantics and its rules can function as sources of redundancy that add to the maximum entropy by enlarging the number of options. The reflexive *animal symbolicum* (Cassirer, 1923) operates by entertaining a symbolic order (Deacon, 1997) that multiplies the envisaged possibilities to interpret events. When the redundancy increases faster than the uncertainty, the latter can be reduced in an evolving system while nevertheless respecting the Second Law (Brooks & Wiley, 1986: 42).

Self-organizing systems use energy to reduce uncertainty internally. Based on Maturana's (1978; cf. Maturana & Varela, 1980) biological theory of self-organization or *autopoiesis*, Luhmann (1986a; [1984] 1995) proposed considering the communication of meaning as the specific operation of the *social* system. From this perspective, the operation of the inter-human communication system is different from but "structurally coupled" with the processing of meaning in individual "consciousness" at the actor level. "Structurally coupled" means that the one system cannot operate without the other as its directly relevant environment: Luhmann's social systems and individual consciousness embed each other reflexively (using "interpenetration"; Luhmann, 1991 and 2002), but the control mechanisms in the two systems are different. Whereas the individual as a unity can be expected to strive for integration and the organization of different meanings in one's mind, society can tolerate functional differentiation and asynchronicity among subsystems of meaning processing with other degrees of freedom. Political discourse, for example, can be coupled to scholarly discourses to varying extents, but the two communication circuits do not have to be updated by each other at each step.

In accordance with the sociological and (most of the) cybernetic tradition, Luhmann ([1984: 102 ff.] 1995: 82f.) defined "information" as the reduction of complexity for a selecting (e.g., observing) system, with explicit reference to Bateson's (1972) definition of information as "a difference that makes a difference." The communication of meaning is thus uncoupled from measurement in terms of Shannon-type information. Shannon's theory assumes that only information *defined* as uncertainty can be communicated, whereas Luhmann assumes that only meaning can be communicated: the inter-human communication system is *defined* as communicating meaning. Consequently, Luhmann concluded that "the most important consequence of this analysis is *that communication cannot be observed, only inferred*" (Luhmann, 1995, at p. 164; italics in the original). Thus, this approach tends to lose the operational perspective of measuring communications in terms of bits of information.

**The research problem**

In this study, we try to solve the above sketched impasse by distinguishing between the communication of information as a relational operation and the positioning of recursive loops that can be generated within communication networks when the information is provided with meaning (Leydesdorff, 2012b). In our opinion, the "communication of meaning" cannot be observed as historical because only Shannon-type information can be communicated in the (measurable) engineering box. However, the processing of meaning among reflexive agents can be considered as adding Weaver's semantic box to the communication as a feedback arrow and thus under conditions expected to reduce the prevailing uncertainty by increasing the maximum entropy.

In other words, this reduction of the uncertainty finds its origin in the increase of the number of options ($N$) when codes can operate upon one another, and accordingly the decrease of all probabilities expressed as relative frequencies ($f/N$). The different codes can be considered as generating different "alphabets" of which the character sets can be cross-tabled. The number of options is then multiplied. The processing of meaning can thus leave an imprint as a feedback by reducing the relative uncertainty or—in other words—increasing the redundancy in the entire configuration.

The Shannon-type mutual information can first be calculated in two dimensions. If one extends the formulas to three or more dimensions, however, the resulting value can no longer be considered as (Shannon-type) information—that is, uncertainty. We shall argue that this value can be considered as a the effect of an increase of the redundancy. Mutual redundancy can be considered as the positional counterpart of the generation of uncertainty in relational communications. In the three-dimensional case (e.g., a Triple Helix of university-industry-government relations), mutual redundancy is equal to mutual information, but when the dimensionality is even, the sign is different. We generalize to the measurement in *N* dimensions. This approach allows us to extend Shannon's mathematical theory of informational communications towards the measurement of exchanges of meaning in terms of different positions. Different positions imply different angles for the reflection of the relational events.

**A *signed* information measure**

Let us proceed to the formalization. The measurement of a negative imprint in terms of bits of information requires a *signed* information measure. As noted above, all Shannon-measures are necessarily positive because of the coupling to the Second Law of thermodynamics. From this perspective, however, one seemingly inconsistent consequence of Shannon's formulas has remained in the case of mutual information in three or more dimensions, because the latter can be negative (e.g., McGill, 1954; Yeung, 2008, at pp. 59f.). Garner & McGill (1956) compared mutual information in three dimensions with the possibility of negative variance in the case of three sources of variance.

The mutual information or transmission (*T*) between *two* variables *x* and *y* is defined (by Shannon) as the difference between the uncertainty in a probability distribution of a discrete random variable *x* (that is, $H_x = -\sum_x p_x * \log_2 p_x$) and the uncertainty in the conditional probability distribution of *x* given *y* ($H_{x|y} = -\sum_{x|y} p_{x|y} * \log_2 p_{x|y}$). In formula format:

$$T_{xy} = H_x - H_{x|y} \tag{1}$$

or equivalently:

$$T_{xy} = H_x + H_y - H_{xy} \tag{2}$$

Rewriting this formula as: $H_{xy} = H_x + H_y - T_{xy}$ makes possible a graphic and set-theoretical interpretation using Figure 2:

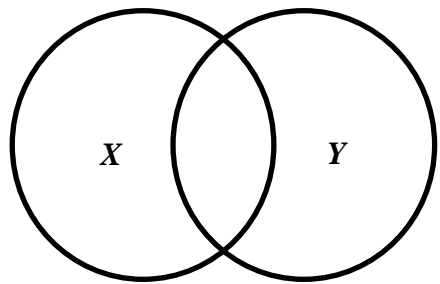

**Figure 2**: Overlapping uncertainties in *x* and *y*.

The total uncertainty in *x* and *y* ($H_{xy}$) is equal to the uncertainty in *x* plus the one in *y*; but one has to subtract the uncertainty in the overlap because this uncertainty would otherwise be counted twice. The uncertainty in the overlap can also be considered as mutual information or transmission ($T_{xy}$).

It follows from this (e.g., Abramson, 1963, at p. 129) that mutual information in the case of *three* variables *x*, *y*, and *z*, or, more generally, 1, 2, and 3 is:[2]

$$T_{123} = H_1 + H_2 + H_3 - H_{12} - H_{13} - H_{23} + H_{123} \tag{3}$$

Using a graphic representation analogous to Figure 2, one can reason as follows using the additive nature of the entropy function in Figure 3:

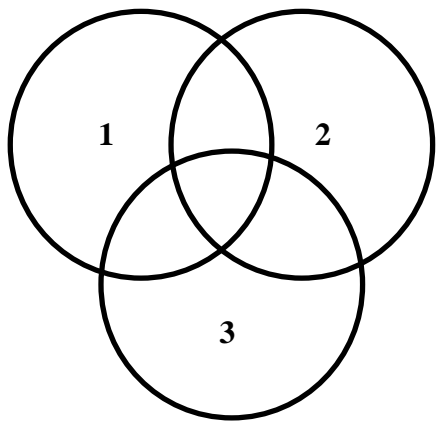

**Figure 3**: Overlapping uncertainties in three variables 1, 2, and 3.

---

[2] The notation is now changed to 1,2,3, ...etc. in order to allow for further extensions beyond three (*1, 2, 3, ...., N*). Following Theil (1972), we generalize the notion of entropy as the average amount of information in a set of symbols to entropy statistics that can be used more abstractly for the analysis of relations among variables.

If one subtracts each of the three mutually overlapping areas in Figure 3 once, the total uncertainty ($H_{123}$) has to be corrected first for each overlap because of double counting, but then the three-way intersection is three times subtracted and consequently no longer included. Therefore, this term ($T_{123}$) has to be added after the subtractions. The total uncertainty is then:

$$H_{123} = H_1 + H_2 + H_3 - T_{12} - T_{13} - T_{23} + T_{123} \tag{4}$$

As in Eq. 2:

$$T_{12} = H_1 + H_2 - H_{12}, \tag{5}$$

and $T_{13}$, $T_{23}$, *mutatis mutandis*.

The substitution of Eq. 5 into Eq. 4 leads to:

$$H_{123} = H_1 + H_2 + H_3 - (H_1 + H_2 - H_{12}) - (H_1 + H_3 - H_{13}) - (H_2 + H_3 - H_{23}) + T_{123} \tag{6}$$

$$T_{123} = H_{123} - H_1 - H_2 - H_3 + (H_1 + H_2 - H_{12}) + (H_1 + H_3 - H_{13}) + (H_2 + H_3 - H_{23})$$
$$= H_1 + H_2 + H_3 - H_{12} - H_{13} - H_{23} + H_{123} \tag{7}$$

The result of Eq. 7 is identical to Eq. 3, but the derivation is different since expressed in terms of the sets and subsets of Figure 3.

Analogously, it follows for higher dimensionality (e.g., four):

$$T_{1234} = H_1 + H_2 + H_3 + H_4 + H_1 - H_{12} - H_{13} - H_{14} - H_{23} - H_{24} - H_{34}$$
$$+ H_{123} + H_{124} + H_{134} + H_{234} - H_{1234} \qquad (8)$$

Note that the sign alternates each time for the lastly added (that is, highest-order) term. However, the different signs *within* each equation (as in Eqs. 7 and 8) provide the opportunity to distinguish between positive and negative contributions to the uncertainty by the mutual information in more than two dimensions.

Building on Ashby (1969), Krippendorff (2009b, at p. 670) provided a general *notation* for this alteration with changing dimensionality—but with the opposite sign (which further complicate the issue; cf. Leydesdorff, 2010a:68)—as follows:

$$Q(\Gamma) = \sum_{X \subseteq \Gamma}(-1)^{1+|\Gamma|-|X|} H(X) \qquad (9)$$

In this equation, $\Gamma$ is the set of variables of which $X$ is a subset, and $H(X)$ is the uncertainty of the distribution; $|\Gamma|$ is the cardinality of $\Gamma$, and $|X|$ the cardinality of $X$.[3]

---

[3] This formula for one, two and three dimensions takes the form (Krippendorff, 1980, 2009a and b):

$$Q(A) = -H(A)$$
$$Q(AB) = H(A) + H(B) - H(AB)$$
$$Q(ABC) = -H(A) - H(B) - H(C) + H(AB) + H(AC) + H(BC) - H(ABC) - H(ABC$$

With the same enthusiasm which Krippendorff (2009a) reports about Ashby (1969), we embraced this potentially negative sign in the mutual information in three dimensions as an indicator of potential reduction of uncertainty in Triple-Helix configurations once it was brought to our attention by Robert Ulanowicz, who had used the same indicator in the context of his ascendancy theory in mathematical biology (Ulanowicz, 1986: 143). This same indicator is used across disciplines (see for an overview: Jakulin, 2005) and sometimes called "configurational information," but it has remained controversial because it is poorly understood. As noted, a signed information measure cannot be interpreted in Shannon's information theory, whereas alternative frameworks for its appreciation have remained ill-defined (Krippendorff, 1980, 2009a and b).

In general, the interaction between two variables can be conditioned by a third source of variation with which both variables can interact in terms of partial and/or spurious correlations. Spurious correlation can reduce uncertainty without being visible in the data without further analysis—and therefore latent (Strand & Leydesdorff, in press). Partial correlation can be measured using conditions in the equation, as with Pearson's $r_{xy|z}$ (Sun and Negishi, 2010) or, non-parametrically, Shannon's $H_{xy|z}$ as a measure of uncertainty in two dimensions conditioned by a third, and the corresponding mutual information $T_{xy|z}$.

---

This measure $Q$ alternates in sign (for dimensions of two or more), but with the opposite sign compared to the mutual information $T$ as used above (Leydesdorff, 2010a).

A spurious correlation can reduce uncertainty, as when two parents reduce uncertainty for a child by providing mutually consistent answers to questions. The answers by one parent inform the child in this case about the expected answers of the other because these answers are coordinated at a systems level. The marriage can be considered as a latent construct in which the child plays a role without constituting it. Note that the reduction of uncertainty cannot be attributed to specific agents, but is generated in the network of relations as a configuration. In the case of divorce, for example, uncertainty among the answers may begin to prevail over the synergy measured as reduction of uncertainty at the system's level.

Similarly, in university-industry-government relations, a strong existing relation between two of the partners can make a difference for the third (Burt, 2001). The Triple-Helix thesis, however, invites opening up to a fourth, fifth, and next-order helical model (Carayannis & Campbell, 2009, 2010; Leydesdorff, 2012a; Leydesdorff & Etzkowitz, 2003). More recently, Ye, Yu, & Leydesdorff (in press) have argued that globalization can unravel the Triple Helix in terms of the fourth dimension of local (e.g., national) versus global organization. Leydesdorff & Sun (2009), for example, elaborated co-authorship data with institutional addresses in terms of industrial, academic, or governmental authors, with the additional dimension national/international, and concluded that the unraveling of this system in Japan was counteracted by retaining reduction of uncertainty from international collaborations. However, a similar effect could not be found in Canadian or Korean data (Kwon *et al*., 2012).

One problem with the extension of mutual information into three or more dimensions, in our opinion, has remained the alternating signs of the equations. What are the consequences in empirical research when three positive terms are added to Eq. 7 in order to obtain Eq. 8, but only a single one ($H_{1234}$) is subtracted? The likelihood of a sign change because of these definitions with alternating signs is considerable.

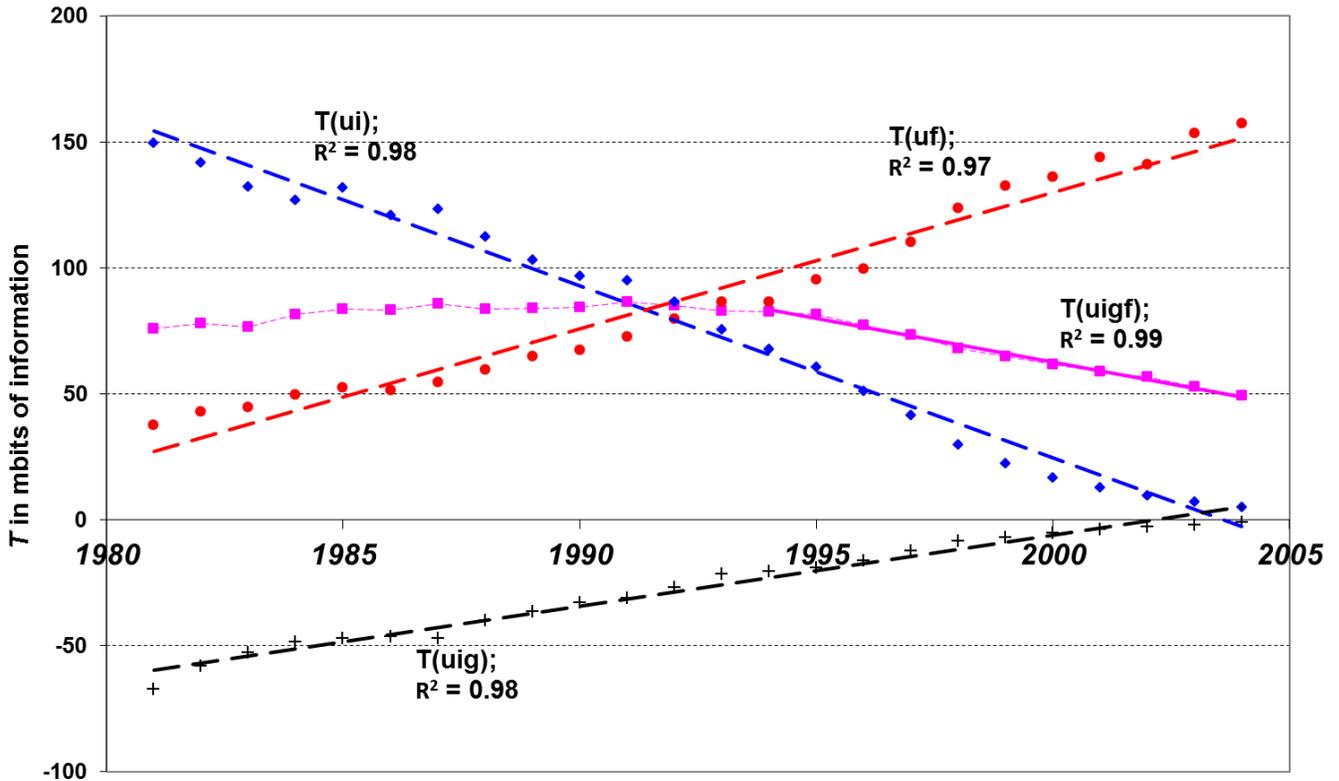

**Figure 4**: Development of mutual relations between authorship with (u)niversity, (i)ndustry, (g)overnmental, and (f)oreign addresses in Japan, 1980-2005; Web-of-Science data. (Source: Leydesdorff & Sun, 2009, at p. 783.)

Figure 4 shows this problem empirically using the data noted above about university-industry-government ("uig") co-authorship relations in Japan combined with international co-authorship relations (using the "f" of "foreign" for an estimate of the non-Japanese addresses involved). Figure 4 shows first that over the entire period 1980-2005, university-industry relations (measured as $T_{ui}$) in two dimensions declined, whereas international (university-foreign) relations (measured as $T_{uf}$) increased steadily. (Note that transmission can also be considered as a measure of co-variation.)

The synergy in the national system of triple-helix relations ($T_{uig}$) thus eroded, but four-dimensional $T_{uigf}$ also exhibited a systematic trend since the early 1990s, that is, after the demise of the Soviet Union and the opening of China. Such a global change in the systems dynamics since about 1993 was found also in comparable studies (e.g., Lengyel & Leydesdorff, 2011; Ye *et al.*, in press) and has been interpreted as a systems change because of further "globalization" emerging in the knowledge-based economies after the end of the Cold War (Leydesdorff & Zawdie, 2010).

The remaining question, in our opinion, is whether this possible decline in $T_{uigf}$ (in Figure 4) since the early 90s indicates a decrease or an increase in uncertainty? Having no other theoretical means available, the authors at the time opted for a decline, but the sign change between the three- and four-dimensional information measure remained worrisome, and this worry became more acute after Krippendorff's (2009a and b) critique (see also Leydesdorff, 2009, 2010a). Krippendorff (2009b) attributed the erratic behavior of the measure to next-order loops that generate redundancy, whereas the transmission values in more than two dimensions were understood as a difference between redundancy generated in next-order loops and Shannon-type information processing generating probabilistic entropy in the interactions.

In our opinion, Krippendorff's specification of mutual information in three dimensions as a difference did not address the conceptual problem of the sign-change when the dimensionality is augmented. In what follows, we propose an approach *within* the framework of the mathematical theory of communication that answers this question and further clarifies that meaning cannot be communicated in the framework of Shannon's theory, but that the circulation of different meanings as a second-order effect of the communication of information may increase the redundancy. The elegance of the new measure of meaning-exchange is that it does not affect the sign in the three-dimensional case, but harmonizes all higher-dimensional cases in accordance with this interpretation.

**The generation of redundancy**

The mutual transmission in the communication of information reduces the uncertainty that prevails using Eqs. 1 and 2 above; but let us assume that new options can be added if meaning processing is additionally involved. Expansion of the number of options ($N$) adds to the maximum entropy: $H_{max} = \log(N)$. Given a prevailing uncertainty $H$, the redundancy $R$ expands since the latter is defined as the fraction of the capacity that is not used for the uncertainty that prevails. In formula format:

$$R = 1 - \frac{H}{H_{max}}$$
$$= \frac{H_{max} - H}{H_{max}}$$

(10)

We focus on the numerator in Eq. 10. Redundancy is then the complement of relative uncertainty: the two add up to the maximum (100% of) entropy. This technical definition of redundancy may seem as counterintuitive as Shannon's definition of "information" since redundancy is intuitively associated with the overlapping information in consecutive messages.

Whereas the assumed increase in redundancy is generated in relation to the information exchanges, it remains redundancy—analytically different from uncertainty—despite this operational coupling. Because we—as analysts—have no information about the size of the additional redundancy, let us for reasons of parsimony assume that the newly generated redundancy is equal in absolute size to the transmission. We can thus define an "excess" information value $Y_{12}$ —equivalent to $H_{12}$ but with the plus sign since we do not correct for the duplication in the case of redundancies—as follows:

$$Y_1 = H_1$$

$$Y_2 = H_2$$

$$Y_{12} = H_1 + H_1 + T_{12} = H_{12} + 2T_{12} \tag{11}$$

Replacement of $H_{12}$ with $Y_{12}$ into Eq. 2 gives analogously the formula for the mutual redundancy $R_{12}$ as follows:

$$R_{12} = H_1 + H_2 - Y_{12} \tag{12}$$

and it follows (using Eq. 11) that:

$$R_{12} = H_1 + H_2 - (H_{12} + 2T_{12})$$

$$= H_1 + H_2 - ([H_1 + H_2 - T_{12}] + 2T_{12}) = -T_{12} \qquad (13)$$

Since $T_{12}$ is necessarily positive (Theil, 1972: 59 ff.), it follows from Eq. 13 that $R_{12}$ is negative and *therefore* cannot be anything else than the consequence of an increased redundancy. This reduction of the uncertainty is measured in bits of information—in other words, as uncertainty—and therefore the sign is negative as a subtraction (Leydesdorff, 2010a: 68). In other words, our mathematical assumption leads to a mutual redundancy generated at the next-order level equal to the transmission in the communication, but with the opposite sign. Consequently, $R_{12}$ can be expressed in terms of negative amounts (e.g., bits) of information, that is, as reduction of uncertainty.

Interestingly, the formulation in Eq. 11 is precisely equivalent to working with "pure sets" (e.g., as in the case of the document sets in Figure 2). One no longer corrects for the overlap—the mutual information—but assumes that the two sets influence each other at a next level using another mechanism, namely, in terms of positions. The overlap (in Figure 2) is thus not subtracted, but *counted twice and therefore redundant*. The influencing at this next level among "pure sets" is different from the relational interaction at the network level that generates mutual information. The mutual redundancy can be considered as a consequence of the *positions* of the communicating systems in relation to one another given the possible communication of information between them. Mutual redundancy $R$ measures the surplus of options that are generated when meaning-processing systems communicate in terms of the information exchange.

Without an information exchange, $T_{12}$ is zero and therefore $R_{12} = -T_{12}$ is also zero. The meaning exchange is *instantiated* by the information exchange as the carrying operation but remains a positional potential. The field-metaphor is useful here: the two sets influence each other by radiating, but disturb each other only to the extent that energy is possibly transferred. Otherwise—for example, in the case of large distances—the sets remain irrelevant black-boxes to each other.

For the three-dimensional case and using Figure 3 above, we can define, in addition to the two-dimensional values of $Y$ (in Eq. 11), a three-dimensional value as follows:

$$Y_{123} = H_1 + H_2 + H_3 + T_{12} + T_{13} + T_{23} + T_{123} \tag{14}$$

Information-theoretically (on the basis of Eq. 6), however:

$$H_{123} = H_1 + H_2 + H_3 - T_{12} - T_{13} - T_{23} + T_{123} \tag{15}$$

and therefore it follows that the difference is:

$$Y_{123} - H_{123} = +2T_{12} + 2T_{13} + 2T_{23}$$
$$Y_{123} = H_{123} + 2T_{12} + 2T_{13} + 2T_{23} \tag{16}$$

Using *Y*-values instead of *H* values for joint entropies in Eq. 3, one obtains:

$$R_{123} = H_1 + H_2 + H_3 - (H_{12} + 2T_{12}) - (H_{13} + 2T_{13}) - (H_{23} + 2T_{23}) +$$
$$+(H_{123} + 2T_{12} + 2T_{13} + 2T_{23}) = T_{123} \qquad (17)$$

In the three-dimensional case, the mutual redundancy is identical to the previously studied mutual information in three dimensions. This explains why the mutual information in more than two dimensions can be negative, and therefore cannot be Shannon-type information.

In other words, in the case of measuring the mutual information in three dimensions, one never measured Shannon-type information, but this mutual redundancy among the sets. The possible reduction of uncertainty—if this measure provides a negative amount (e.g., bits) of information—can also be considered as a measure of the *synergy* among three sources of variance when specifically positioned in relation to one another. These sources are positioned in terms of their relations in the networks that carry the systems in terms of operations. The network of relations span an architecture in which all links and nodes have a position.

Etzkowitz & Leydesdorff (2000), for example, have used the metaphor of an overlay of communications or "hypercycle" (Leydesdorff, 1994) in university-industry-government relations. The synergy measure used as a Triple-Helix indicator can alternatively be named "configurational information" or Yeung's (2008) *signed* information measure that was denoted by him with $\mu$. In our opinion, Yeung's $\mu^*$ can be considered as mutual redundancy; if $R_{123} < 0$ in bits (or any other measure) of information, the prevailing uncertainty is reduced. Ulanowicz (2009) elaborates on this measure using the metaphor of auto-catalysis: the third variable can operate as a catalyst on the two others by closing a loop among the three. A spurious correlation can then be generated auto-catalytically.

Let us now extend to the four-dimensional case by drawing Figure 5, as follows:

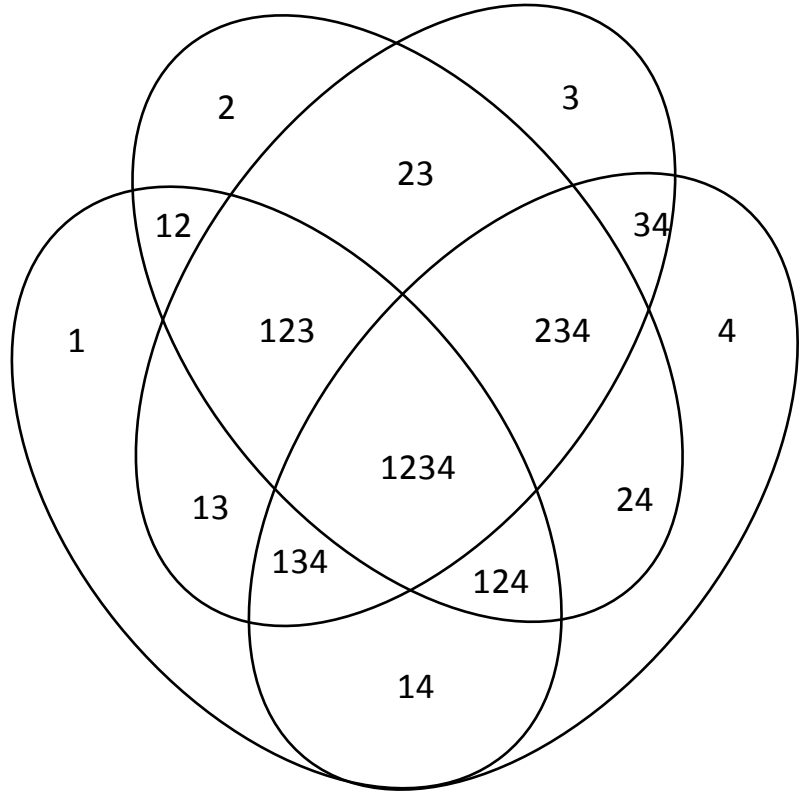

**Figure 5**: Overlapping uncertainties in four variables 1, 2, 3, and 4 (Rousseau, 1998; Durrett, *s.d.*).

We subtract the following two equations:

$$H_{1234} = H_1 + H_2 + H_3 + H_4 - T_{12} - T_{13} - T_{14} - T_{23} - T_{24} - T_{34} + T_{123} + T_{124} + T_{134} + T_{234} - T_{1234} \quad (18)$$

$$Y_{1234} = H_1 + H_2 + H_3 + H_4 + T_{12} + T_{13} + T_{14} + T_{23} + T_{24} + T_{34} + T_{123} + T_{124} +$$

$$T_{134} + T_{234} + T_{1234} \tag{19}$$

with the following result:

$$Y_{1234} = H_{1234} + 2T_{12} + 2T_{13} + 2T_{14} + 2T_{23} + 2T_{24} + 2T_{34} + 2T_{1234} \tag{20}$$

Using $Y$ values instead of $H$ in Eq. 8 for the joint entropies, we obtain (using also Eqs. 11 and 16 for the substitution):

$$\begin{aligned}
R_{1243} &= H_1 + H_2 + H_3 + H_4 - (H_{12} + 2T_{12}) - (H_{13} + 2T_{13}) - (H_{14} + 2T_{14}) - (H_{23} \\
&\quad + 2T_{23}) - (H_{24} + 2T_{24}) - (H_{34} + 2T_{34}) + (H_{123} + 2T_{12} + 2T_{13} + 2T_{23}) \\
&\quad + (H_{124} + 2T_{12} + 2T_{14} + 2T_{24}) + (H_{134} + 2T_{13} + 2T_{14} + 2T_{34}) + (H_{234} \\
&\quad + 2T_{23} + 2T_{24} + 2T_{34}) - (H_{1234} + 2T_{12} + 2T_{13} + 2T_{14} + 2T_{23} + 2T_{24} \\
&\quad + 2T_{34} + 2T_{1234}) \\
&= H_1 + H_2 + H_3 + H_4 - H_{12} - H_{13} - H_{14} - H_{23} - H_{24} - H_{34} + H_{123} \\
&\quad + H_{124} + H_{134} + H_{234} - (H_{1234} + 2T_{1234}) \\
&= T_{1234} - 2T_{1234} \\
&= -T_{1234} \tag{21}
\end{aligned}$$

In summary, we can see that the problem of the change of sign is resolved by using $R$ instead of $T$ since:

$$R_{12} = -T_{12} \tag{22}$$

$$R_{123} = T_{123} \tag{23}$$

$$R_{1234} = -T_{1234} \tag{24}$$

Equations 22, 23, and 24 show first that *R* can have a negative sign; Equation 22 warrants this in the two-dimensional case. Although (Shannon-type) information and redundancy are positive, we can measure mutual redundancy also as reduction of uncertainty. In the three-dimensional or, more generally, the odd-dimensional cases, the mutual information among these dimensions is defined as equal to the measurement of the mutual redundancy, but in the even-dimensional cases the sign of this redundancy is opposite to the mutual information.

In summary, the alternating sign is thus explained by assuming that the mutual positioning of reflexive communication systems in terms of relations can generate mutual redundancy that is (technically) equal in size to the (previously defined) mutual information in the relations. Krippendorff's (2009a) insistence on the interaction information $I_{ABC \to AB:AC:BC}$ as the Shannon-type information in three dimensions different from $T_{ABC}$ can then be fully appreciated. The signed information measure of Yeung $\mu^*$ has always measured a non-Shannon redundancy *R* (albeit with a potential sign-change depending on the dimensionality), whereas Krippendorff's interaction information can be considered a Shannon-type measure.[4]

---

[4] Krippendorff's approximation of this value can be extended to more than three dimensions using Occam v3.3.10 at http://occam.research.pdx.edu/occam/weboccam.cgi?action=search (Willett & Zwick, 2004; cf. Krippendorff, 1986).

**The interpretation**

Our information-theoretical interpretation first follows Krippendorff's (2009b, at p. 676) assertion "that *interactions with loops entail positive or negative redundancies, those without loops do not.*" Mutual information in three or more dimensions was considered by Krippendorff as the difference between redundancy thus generated and the Shannon-type information generated in the interactions. Or, to use Weaver's (1949) terminology, between noise generated in the engineering box and the (next-order) semantic noise-generator.

We follow Weaver's interpretation by considering the two boxes (of engineering noise and semantic noise in Figure 1) as operating in parallel with a potential net result at the level of the network channel. The two operations are analytically independent, but mutual redundancy can only be generated in relations; that is, when the channel operates. Both boxes can be considered as conditions on the information exchange, but mutual redundancy adds to the maximum entropy of the channel. In other words, it changes the channel capacity and is not a Shannon-type information that assumes a fixed channel. This expansion of the maximum entropy may more than compensate for the interaction information that is generated relationally (Brooks & Wiley, 1986; Leydesdorff, 2011). Relational operation is a necessary condition for the generation of mutual redundancy. Note that there can also be other sources of redundancy, for example, through codification of the communication within a system (see below).

Following Weaver (1949), one can assume that meaning-processing can add to the redundancy, but this "semantic noise" had not hitherto been specified. Our proposal is that for reasons of parsimony the mutual redundancy can be mathematically defined as equal to the mutual information in the respective number of dimensions, but with a possible correction for the sign as specified above. However, redundancies do not relate operationally—but only condition the relational operation. From a set-theoretical approach, therefore, the sets were first considered as independent ("pure" or orthogonal). The "pure sets" can be redundant in terms of their information contents and this can structure the relations as "semantic noise" (Figure 1). Whether such structuring adds to or reduces the uncertainty in the the total system of variables is an empirical question.

The choice of this value for and sign of the mutual redundancy is convenient because

1. In the three-dimensional case, $R_{123} = T_{123}$ and thus the *signed* information measure that has been used in empirical studies throughout the literature remains a valid measure, albeit with a different interpretation. If $R_{123} < 0$ uncertainty is reduced, and if $R_{123} > 0$ uncertainty is increased;
2. $R_{12} = -T_{12}$ in the two-dimensional case; $R_{12}$ is necessarily $\leq 0$, since $T_{12} \geq 0$ (Theil, 1972: 59f.); measurement of $R_{12}$ as *reduction* of uncertainty makes mutual redundancy consistent with Shannon's theory;
3. In the *N*-dimensional case, the sign of the mutual information *T* would alternate, but the sign of the mutual redundancy *R* remains consistent;
4. This reasoning does not affect the specification of Krippendorff's "interaction information" ($I_{ABC \rightarrow AB:AC:BC}$) as a Shannon-type information in three or more dimensions; *R* is not a Shannon-type information, but the calculus remains thus consistent with Shannon's mathematical theory of communication.

Let us proceed to a systems-theoretical appreciation. From a Shannon-perspective, meaning cannot be communicated relationally, but only information can be communicated between a sender and a receiver. Bateson's (1972) "difference which makes a difference," however, is always provided with reference to a system. By acknowledging the difference made, the incoming (Shannon-type) uncertainty is *positioned* within a receiving system, and thus provided with local meaning. Unlike information processing that is based on operational coupling in relations, meaning processing operates as a next-order loop in terms of positions that influence one another as fields.

For didactic purposes, the difference between mutual redundancy between systems with different positions and mutual information in their relations can be compared with the difference between the additive and subtractive mixing of colors: red, green, and blue (or cyan, magenta, and yellow) add up to white additively—for example, in terms of three lamps that are positioned differently—but the result of operational mixing of these three colors (relationally) leads to subtractive colour-mixing and ultimately to black (Figure 6).

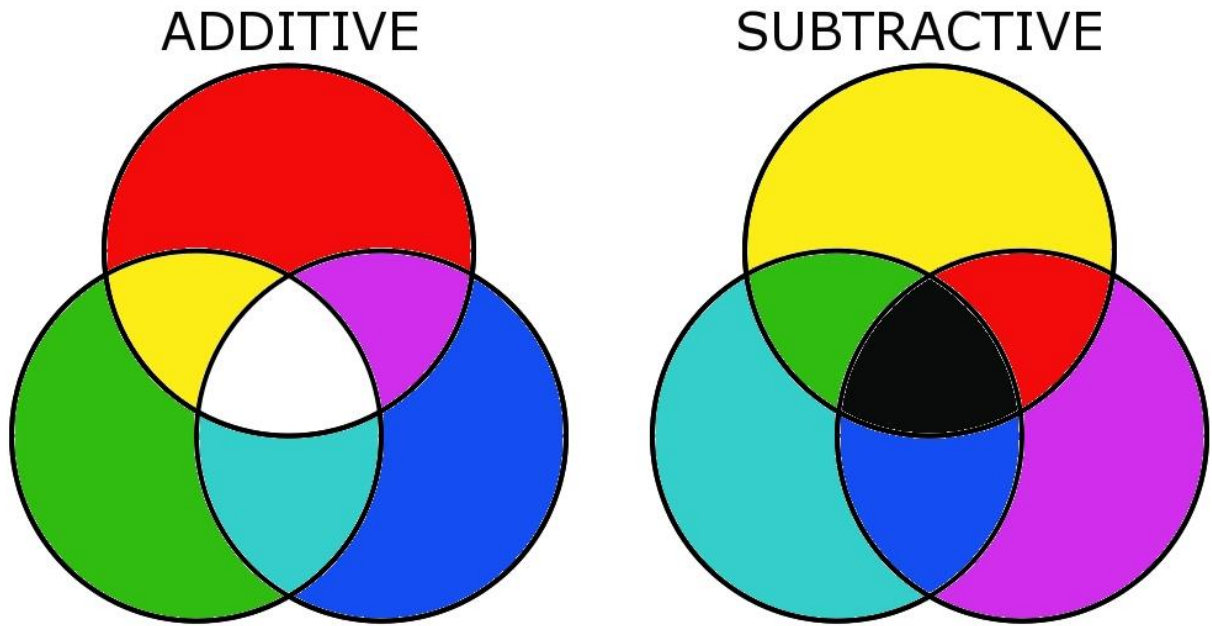

**Figure** 6: Additive and subtractive color mixing.

Certain elements of Luhmann's ([1984], 1995) social-systems theory can also be helpful for explaining this functioning of meaning-processing in terms of positions. Based on Parsons' (1968) theory of the symbolic generalization of media, as in the case of money and power, Luhmann considered functional differentiation among the (symbolically generalized) media to be based on different codes operating in subsystems of communication. For example, he suggested true/false as the code of communication in scientific communication *versus* transactions ("payment") as the code on the market. Parsons (1963a and b) distinguished between "power" and "influence" in these terms, and one can also consider affection ("love") as a specific coding of the communication (Luhmann, 1986b). The codes thus color the communications like colors of light without yet (analytically) implying the necessity of a relational ("operational") coupling or mixing. Without communication, however, the codes cannot operate.

Following Maturana's (e.g., 1978) theory of *autopoiesis*, Luhmann ([1984], 1995; 1992) originally argued that the "function systems"—specifically coded communication (sub)systems—are operationally closed and structurally coupled. This metaphor is (meta-)biological: a functionally differentiated cell in a bodily organ such as the liver cannot perform a lung-cell function once differentiation has taken place (in the *gastrula* stage of embryonic development), but both cells are structurally part of the same body. Functionally differentiated systems operate in parallel, but they can influence each other in terms of operational couplings at the network level (for example, via the bloodstream).

In response to criticism, Luhmann (1997, at p. 788) added that communications can be operationally coupled in events, whereas the same communication can be provided with different meanings at different places. He elaborated the example of a physician who provides an employee with a note that legitimates absence from the workplace because of a health condition. In this event, the health system is operationally coupled to the economic system, but the meanings are updated in each of the different subsystems that are asymmetrically coupled in the operation. One can also consider the coupling as a "translation" or "overflow" (Callon, 1998).

In terms of Triple-Helix relations, for example, patents can be considered as output of R&D systems and inputs to the economy. They serve at the same time intellectual property protection. The three different institutional spheres are recombined in the specific organization of a patent as an instant in the relevant flows. In other words, each relational operation is contingent and historical: mutual information is generated in the interaction that is necessarily *organized* in (and mediated by) historical events. The functional subsystems, however, *self-organize* in terms of updating their information contents at the level of their respective systems using specific codes. New meanings can be generated as possible options or, in other words, redundancies in overlaps. The subsystems are positioned in relation to one another in a function space that is different from the network space in which the operations are performed relationally (Bradshaw & Lienert, 1991; Simon, 1962; cf. Leydesdorff, in press).

Simon (1973, at p. 19 ff.) argued that complex systems can be expected to operate with alphabets. In other words, the codes of communication can be considered as alphabets. However, the number of coding systems can grow (or shrink) over historical times. Parsons' structural-functionalism could thus be historicized (Merton, 1957; Mittroff, 1974). More codes allow for more complexity in the communication by generating more possible redundancies. In each event (or "instantiation"; Giddens, 1979), however, specific codes are (de-)selected more than others, and thus different mixtures of meanings are continuously generated over historical time.

Note that self-organization is an evolutionary dynamics hypothesized at a trans-historical level—"genotypical" in evolutionary terms (Hodgson & Knudsen, 2011)—that can be expected to leave "phenotypical" (that is, observable) imprints on the historical and observable organization of events. The supra-individual level of intentionality, however, remains an order of expectations that can fruitfully be hypothesized, but cannot be observed directly (Leydesdorff, 2012b).

Giddens (1979, at p. 71) expressed this relation between "absent" redundancies and "present" relations so well that his text merits quotation in this context:

> "That is to say, the differences which constitute social systems reflect a dialectic of presences and absences in space and time. But these are only brought into being and reproduced via the virtual order of differences of structures, expressed in the duality of structure. The differences that constitute structures, and are constituted structurally, relate 'part' to 'whole' in the sense in which the utterance of a grammatical sentence presupposes the absent corpus of syntactical rules that constitute the language as a totality. The importance of this relation of moment and totality for social theory cannot be exaggerated, since it involves a dialectic of presence and absence which ties the most minor or trivial forms of social action to structural properties of the overall society (and, logically, to the development of mankind as a whole)."

As Giddens (1979) went on to explain, institutions therefore do not work "behind the backs" of the social actors who produce and reproduce them as structured expectations in interactions.

From a communication-theoretical (as different from Giddens' action-oriented) perspective, Luhmann (1975, 1986c) specified three levels of "structuration": (*i*) interactions at the micro level that can be expected to encompass potentially all functions—the full spectrum—of interhuman communications; (*ii*) aggregates of communication can partially and selectively integrate differently coded communications relationally at specific (historically observable!) interfaces; and (*iii*) at the level of society, events can be appreciated and coded differently depending on the macro-structures that have been developed in inter-human communications as structures of expectations. For example, the rule of law and the functioning of the market—as an invisible hand operating at the supra-individual level—are two such counterfactual coordination mechanisms operating among us in terms of mutual expectations.

When information is communicated between (sub)systems using different codes—that is, the encoding and decoding of the messages are performed with different algorithms—then the received message can be given a variety of meanings, and thus the number of options is rapidly enlarged (Leydesdorff, 1997). The maximum uncertainty of two interacting systems thus can be expected to increase with this redundancy, but not automatically since mutual redundancy is only generated in configurations of relations.

The mutual redundancy adds semantic noise to the engineering noise. Our proposal is to estimate this (potentially negative) contribution to the uncertainty as equal to the mutual information—because one has no other information—but with the opposite sign in the case of an uneven number of dimensions. The advantage of our definition of mutual redundancy in terms of "pure sets" is its consistency with Shannon's (1948) framework and the elaboration of Weaver's (1949) intuitions in this context (Shannon & Weaver, 1949). Furthermore, this extension of the mathematical theory of communication resolves the problem of the alternating sign in the mutual information in more than two dimensions, and can be interpreted in terms of "pure sets" that operate concurrently, and thus can be redundant (since repetitive) from a system's perspective. The redundant signals always form a surplus in the communication. Mutual information in more than two dimensions has hitherto been misunderstood as an information measure; it is actually a measure of mutual redundancy.

**Double contingency**

One should remain aware of the confusing metaphors when mixing the information-theoretical with Luhmann's communication-theoretical perspective. When one uses "information" and "communication" with reference to Shannon's information theory, the concepts are defined without substantive meaning as yet and can be measured, for example, in bits of information. The specification of a system of reference provides the communication with dimensionality and meaning: for example, one can communicate with words or with money. Both financial transactions and co-occurrences of words can be measured in terms of bits of information (Leydesdorff, 1995).

As noted, sociologists (including Luhmann) use "communication" specifically for inter-human communication. In inter-human communication not only information is exchanged, but intentionality can also be expected to play a role (Husserl, 1929). Meaning, however, cannot be communicated directly, but only in terms of instantiations in events where different meanings are related in terms of information exchanges. Yet, meaning in terms of mutual expectations also influences each interhuman communication. If the "semantic noise" were zero, the interhuman communication could not be received as meaningful; it might be coded—and then possibly discarded—as merely engineering noise.

Using different terminologies, 20$^{th}$-century sociologists and phenomenologists have tried to express this duality between uncertainty and meaning processing in inter-human communications. For example, Mead (1934) distinguished between "*I*" and "*me*"—the individual being at the same time both subject and object; Husserl (1929) returned to Descartes' *res extensa*" and "*res cogitans*," where the latter remains in the domain of intentionality; Geertz' (1973) difference between "emic" and "etic" in anthropology refers to whether the analyst takes both dimensions into account by "understanding" or merely observes behaviour; and, as noted, Giddens (1979) used "the duality of structure," but he did not wish to specify the operation of this duality otherwise than methodologically (e.g., Giddens, 1984: 304; cf. Bryant & Jary, 1991: 96; Leydesdorff, 1993: 54f.).[5] As noted above, Luhmann specified "structuration" in terms of interaction, organization, and self-organization.

Luhmann also elaborated on Parsons' notion of "double contingency." Parsons (1968) distinguished two analytically different contingencies as specific for inter-human relations: one in terms of the *res extensa* and one in terms of what the contingent other(s)—things, people—mean to us. The second of these contingencies generates what he called a "double contingency" in the inter-human encounter: *Ego* expects *Alter* to entertain expectations similar to *Ego*'s own expectations (Luhmann 1995: 103ff; Parsons 1951: 91ff, 1968). The sharing and exchanging of expectations in a double contingency opens "horizons of meaning" (Husserl, 1929, 1962; Leydesdorff, 2012b).

In our opinion, the exchange of meaning remains an epi-phenomenon that accompanies the exchange of information in human language or by way of symbolically generalized media of communication for the non-verbal exchange as a second-order effect. However, the epi-phenomenal level, while constructed and continuously reconstructed bottom-up, can take over control when sufficiently codified, by providing references to "horizons of meaning." The feedback can then also become a feed-forward to other possible meanings because redundancy (the number of options) can be proliferated faster than (Shannon-type) information, for example, in knowledge-based systems. The extent to which this inversion happens remains an empirical question. The measurement of mutual redundancy makes such questions amenable to empirical research (e.g., Leydesdorff & Strand, in press).

---

[5] Giddens (1976: 162) specified also "double hermeneutics" as constitutive for the social sciences. However, the "double hermeneutics" refers specifically to the two-way communication between scholarly and lay concepts in the social sciences, whereas "double contingency" plays a role in all inter-human communication because of the reflexivity involved.

**Discussion**

The definition of the "mutual redundancy" as uncertainty (or reduction of uncertainty), but equal in the absolute size to the "mutual information" may seem to impose an arbitrary limitation on the number of options that can be added to the maximum entropy in next-order loops of meaning processing. However, this limitation makes our reasoning consistent with the mathematical theory of communication and provides the mutual information in three dimensions with an interpretation, whereas such an interpretation has been unavailable hitherto (despite the many empirical applications of the measure). The mutual information in three dimensions $\mu^*$, however, *was never Shannon-type information but mutual redundancy $R_{123}$*. The Shannon-type interaction information $I_{ABC \rightarrow AB:AC:BC}$ in three dimensions obeys a very different calculus (Krippendorff, 1982, 2009a and b).

The lower limit of mutual redundancy in two dimensions ($R_{12}$) is zero if the mutual information is zero. This is unproblematic; but an upper limit of $R_{12} = -T_{12}$ seems unnecessarily constraining on meaning processing while meanings may develop much faster in terms of reflexive options than can be retained historically. Wishes, expectations, and phantasies, for example, proliferate faster than actions (Weinstein & Platt, 1969). However, this proliferation is *within* systems and not in terms of their operational relations. Mutual redundancy is only a part of the total redundancy of interacting communication systems.

Let us combine Luhmann's (1995) and Simon's (1973) conjectures that inter-human communications operate with an alphabet of functionally different codes. If *N* codes are used for the communication—instead of a single set of codes—mutual redundancies can be generated concurrently in all of these relations as parallel channels of the communication. Already at the level of language—that is, below the order of symbolically generalized codes of communication—the reflexive observer and the participant are able to change roles, and thus mutual contingencies can rapidly be instantiated. The option to change repertoires in negotiations in order to enlarge the number of options is part of the communicative competencies of participants, for example, in the construction of discourse coalitions (Hajer *et al*., 1993; Levidow, 2009; Wagner & Wittrock, 1990).

Within each of Luhmann's "function systems" or Giddens' "sets of rules and resources" (e.g., science, economy or polity), symbolic generalization of the code adds another degree of freedom to the processing of meaning. For example, whereas economists and engineers may provide different meanings and values to patents, their lawyers can be expected to use a very different rationale when litigating in the court room. More options for providing meaning are thus generated by functional differentiation of the codes of communication, and thus redundancy may proliferate more rapidly than only in terms of mutual redundancies. These longer-term developments over time stand in orthogonal relation to the interaction information and mutual redundancy generated in instantiations (or events) at specific moments in time.

Note that symbolic generalization can also be expected to have a next-order effect on coding as an operation. Whereas (undifferentiated) meaning is first provided from the perspective of hindsight—that is, "incursively" instead of "recursively"—symbolic generalization can lead to hyper-incursivity. The calculus of meaning processing can thus be embedded in the theory and computation of anticipatory systems (Rosen, 1985; Dubois, 1998 and 2003); but such an extension would lead us beyond the scope of the present study (cf. Leydesdorff, 2010b).

The message to remember is that mutual redundancy is not total or maximum redundancy, just as mutual information is only part of a prevailing uncertainty. From this perspective, the present study can be characterized as just a first step towards a calculus of meaning, that is, the specification of redundancies and uncertainties in the processing of meaning. We here elaborated mutual redundancy from a sociological perspective—because the understanding of "meaning" is urgent at this level—but one should note that synergy can also be measured using $T_{xyz}$ ($= R_{xyz}$) at many other systems levels.

For example, Maturana (1978, at p. 49) formulated (albeit with possibly another intention) as follows: "(I)f an organism is observed in its operation with a second-order consensual domain, it appears to the observer as if its nervous system interacted with internal representations of the circumstances of its interactions, and as if the changes of state of the organism were determined by the semantic value of these representations." Maturana then argued that one should not in the biological case be misled by the linguistic metaphor. Our specific research question in this study, however, was about the generation of redundancy in inter-human communication; that is, within a semantic domain where second-order representations function in the communication. The further extension to a biological (or other) system that can be considered also as a semantic domain is expected to require a different approach.

**Conclusions and summary**

This study originated from the puzzle caused by the possible change in sign when mutual information is measured in a next dimension. In the discussions about extending the number of helices beyond a Triple Helix of university-industry-government relations to a quadruple, quintuple, or n-tuple set of helices, this problem had urgently to be resolved (Carayannis & Campbell, 2009 and 2010; Leydesdorff, 2012a; Leydesdorff & Sun, 2009; Kwon *et al.*, 2012). We could solve the technical puzzle by considering "pure" sets, that is, without accounting for the overlap as transmission, but by adding "excess" entropy to the equation as additional redundancy. The concept of "mutual redundancy" then had to be provided with an interpretation. We have argued that mutual information in three dimensions as defined and used in the literature since its invention by McGill (1954) has always been a measurement of mutual redundancy. If $R_{123} < 0$, this mutual redundancy has a negative effect on the uncertainty that prevails.

When information is communicated between reflexive (sub)systems using different codes—that is, the encoding and decoding of the messages are performed with different algorithms—the received message can be given a variety of meanings and thus the number of options is enlarged. The two interacting systems can then be expected to generate mutual redundancy, but not automatically. Mutual redundancy is generated in relations. We propose to estimate this uncertainty as equal to the mutual information—because one has no other information—but with a consistent (and therefore sometimes opposite) sign. The correct sign (when uncertainty is expressed in bits of information) follows from our definitions. The advantage of this proposal is its consistency with the Shannon framework in which probabilistic entropy is always positive and increasing. Loops of communication could hitherto not be appreciated from this perspective. Furthermore, this extension of the mathematical theory of communication solves the problem of the alternating sign in mutual information in more than two dimensions by clarifying this concept as mutual redundancy.

We provide a systems-theoretical appreciation in terms of Luhmann's (1995) social-systems theory and Giddens' (1979) structuration theory. Luhmann's scheme first distinguished between human agents who process meaning consciously and inter-human communication. These two systems are structurally coupled, but unlike the biological concept, operational closure is not achieved because meaning is provided reflexively, and thus mutual redundancy is also generated in the *operation* of "interpenetration" at the reflexive level (Luhmann, 1991 and 2002).[6]

We have argued that Luhmann's double structure of organization and self-organization is not very different from the structures of Giddens (1979: 71 and 1984: 377) as different "sets of rule-resource relations" that are "instantiated"—and therefore interfaced—in actions as events. If one changes Giddens' "structuration of action" into the "structuration of expectations," the two theories become virtually identical (Leydesdorff, 2010b). Reflexive (inter-)actions (by and among human beings) can then be considered as specific ways to organize expectations. It seems to us that our proposal precisely fills the duality between "absent" self-organization and "present" organization/instantiation in terms that can also be measured in bits of information.

---

[6] Within the social system, such a duality between (internal) structural coupling (Luhmann, 1993, Ch. 10) and operational coupling was also specified (Luhmann, 1995, at p. 778) among the subsystems when functionally differentiated in terms of the symbolic generalization of codes of communication. We used above the coupling between the economic and the health system by a physician's note following his own example.

**Aknowledgements**
We are grateful to Wouter de Nooy, Rob Hagendijk, Ronald Rousseau, and two anonymous referees for comments on a previous version of this manuscript.

Leydesdorff, L., & Sun, Y. (2009). National and International Dimensions of the Triple Helix in Japan: University-Industry-Government versus International Co-Authorship Relations. *Journal of the American Society for Information Science and Technology 60*(4), 778-788.

Leydesdorff, L., & Zawdie, G. (2010). The Triple Helix Perspective of Innovation Systems. *Technology Analysis & Strategic Management, 22*(7), 789-804.

Leydesdorff, L., & Strand, Ø. (in press). The Swedish System of Innovation: Regional Synergies in a Knowledge-Based Economy. *Journal of the American Society for Information Science and Technology.*

Luhmann, N. (1975). Interaktion, Organisation, Gesellschaft: Anwendungen der Systemtheorie. In M. Gerhardt (Ed.), *Die Zukunft der Philosophie* (pp. 85-107). München: List.

Luhmann, N. (1984). *Soziale Systeme. Grundriß einer allgemeinen Theorie*. Frankfurt a. M.: Suhrkamp.

Luhmann, N. (1986a). The autopoiesis of social systems. In F. Geyer & J. v. d. Zouwen (Eds.), *Sociocybernetic Paradoxes* (pp. 172-192). London: Sage.

Luhmann, N. (1986b). *Love as Passion: The Codification of Intimacy*. Stanford, CA: Stanford University Press.

Luhmann, N. (1986c). Intersubjektivität oder Kommunikation: Unterschiedliche Ausgangspunkte soziologischer Theoriebildung. *Archivio di Filosofia, 54*(1-3), 41-60.

Luhmann, N. (1991). Die Form "Person." *Soziale Welt, 42*(2), 166-175.

Luhmann, N. (1993). *Das recht der Gesellschaft*. Frankfurt a.M.: Suhrkamp.

Luhmann, N. (1995). *Social Systems*. Stanford, CA: Stanford University Press.

Luhmann, N. (2002). How Can the Mind Participate in Communication? In W. Rasch (Ed.), *Theories of Distinction: redescribing the Descriptions of Modernity* (pp. 169–184). Stanford, CA: Stanford University Press.

Maturana, H. R. (1978). Biology of language: the epistemology of reality. In G. A. Miller & E. Lenneberg (Eds.), *Psychology and Biology of Language and Thought. Essays in Honor of Eric Lenneberg* (pp. 27-63). New York: Academic Press.

Maturana, H. R., & Varela, F. (1980). *Autopoiesis and Cognition: The realization of the Living*. Boston: Reidel.

McGill, W. J. (1954). Multivariate information transmission. *Psychometrika, 19*(2), 97-116.

Mead, G. H. (1934). The Point of View of Social Behaviourism. In C. H. Morris (Ed.), *Mind, Self, & Society from the Standpoint of a Social Behaviourist. Works of G. H. Mead* (Vol. 1, pp. 1-41). Chicago and London: University of Chicago Press.

Merton, R. K. (1957). *Social theory and social structure, rev. ed*. Glencoe, IL: The Free Press.

Miller, G. A. (1951). *Language and communication*. New York, NY: McGraww-Hill.

Mitroff, I. I. (1974). *The subjective side of science*. Amsterdam: Elsevier.

Parsons, T. (1951). *The Social System*. New York: The Free Press.

Parsons, T. (1963a). On the Concept of Political Power. *Proceedings of the American Philosophical Society, 107*(3), 232-262.

Parsons, T. (1963b). On the Concept of Influence. *Public Opinion Quarterly 27 (Spring)*, 37-62.

Parsons, T. (1968). Interaction: I. Social Interaction. In D. L. Sills (Ed.), *The International Encyclopedia of the Social Sciences* (Vol. 7, pp. 429-441). New York: McGraw-Hill.

Parsons, T., & Shils, E. A. (1951). *Toward a General Theory of Action*. New York: Harper and Row.

Rosen, R. (1985). *Anticipatory Systems: Philosophical, mathematical and methodological foundations*. Oxford, etc.: Pergamon Press.
43